\input harvmac
\input epsf

\overfullrule=0pt
\abovedisplayskip=12pt plus 3pt minus 3pt
\belowdisplayskip=12pt plus 3pt minus 3pt
%
\def\tilde{\widetilde}

\def\cN{{\cal N}}

\def\bns{B_{NS,NS}}
\def\brr{B_{RR}}

\font\zfont = cmss10 
\font\litfont = cmr6

\def\bigone{\hbox{1\kern -.23em {\rm l}}}
\def\ZZ{\hbox{\zfont Z\kern-.4emZ}}
\def\half{{\litfont {1 \over 2}}}


\lref\klebwit{I. Klebanov and E. Witten, {\it ``Superconformal Field 
Theory on Threebranes at a Calabi- Yau Singularity''}, Nucl.Phys.
{\bf B536} (1998) 199; hep-th/9807080.}
\lref\malda{J. Maldacena, {\it ``The Large N Limit of
Superconformal Field Theories and Supergravity''},
Adv. Theor. Math. Phys. {\bf 2} (1998) 231; hep-th/9711200.}
\lref\gubkleb{S. S. Gubser and I. R. Klebanov, {\it ``Baryons and Domain
Walls in an $\cN =1$ Superconformal Gauge Theory''},
Phys. Rev. {\bf D58} (1998) 125025; hep-th/9808075.}
\lref\gjm{D. Ghoshal, D. Jatkar and S. Mukhi, {\it ``Kleinian 
Singularities and the Ground Ring of $c=1$ String Theory''},
Nucl. Phys. {\bf B395} (1993), 144; hep-th/9206080.}
\lref\bsv{M. Bershadsky, V. Sadov and C. Vafa, {\it ``D-strings on
D-manifolds''}, Nucl. Phys. {\bf B463} (1996) 398; hep-th/9510225.}
\lref\oogvaf{H. Ooguri and C. Vafa, {\it ``Two-Dimensional Black Hole and
Singularities of CY Manifolds''}, Nucl. Phys. {\bf B463} (1996) 55;
hep-th/9511164.}
\lref\stromconif{A. Strominger, {\it ``Massless Black Holes and
Conifolds in String Theory''}, Nucl. Phys. {\bf B451} (1995) 96;
hep-th/9504090.}
\lref\dougmoore{M. Douglas and G. Moore, {\it ``D Branes, Quivers and ALE
Instantons''}, hep-th/9603167.}
\lref\dgm{M. Douglas, B. Greene and D. Morrison, {\it ``Orbifold
Resolution by D-Branes''}, Nucl. Phys. {\bf B506} (1997) 84;
hep-th/9704151.} 
\lref\fractional{J. Polchinski, {\it ``Tensors from K3 Orientifolds''},
Phys. Rev. {\bf D55} (1997) 6423; hep-th/9606165\semi
M. Douglas, {\it ``Enhanced Gauge Symmetry in M(atrix)
Theory''}, JHEP {\bf 07} (1997) 004; hep-th/9612126\semi 
D.-E. Diaconescu, M. Douglas and J. Gomis, {\it ``Fractional Branes and 
Wrapped Branes''}, JHEP {\bf 02} (1998) 013; hep-th/9712230.}
\lref\acl{B. Andreas, G. Curio and D. L\"ust, {\it ``The Neveu-Schwarz 
Five-Brane and its Dual Geometries''}, JHEP {\bf 10} (1998) 022; 
hep-th/9807008.}
\lref\kls{A. Karch, D. L\"ust and D. Smith, {\it ``Equivalence of 
Geometric Engineering and Hanany-Witten via Fractional Branes''},
Nucl. Phys. {\bf B533} (1998) 348; hep-th/9803232.} 
\lref\gns{S. Gubser, N. Nekrasov and S. Shatashvili, {\it
``Generalized Conifolds and Four Dimensional $\cN=1$ Superconformal
Theories''}; hep-th/9811230.}
\lref\hanwit{A. Hanany and E. Witten, {\it ``Type IIB Superstrings, BPS
Monopoles, and Three-Dimensional Gauge Dynamics''}, Nucl. Phys. {\bf 
B492} (1997) 152; hep-th/9611230.}
\lref\witqcd{E. Witten, {\it ``Branes and the Dynamics of QCD''},
Nucl. Phys. {\bf B507} (1997) 658; hep-th/9706109. }
\lref\candelas{P. Candelas and X. C. de la Ossa, {\it ``Comments on 
Conifolds''},  Nucl. Phys. {\bf B342} (1990) 246.}
\lref\mori{D. Morrison and R. Plesser, {\it ``Non-Spherical
Horizons--I''}; hep-th/9810201.}
\lref\angel{A. Uranga, {\it ``Brane configurations for Branes at
Conifolds''}, JHEP {\bf 01} (1999) 022; hep-th/9811004.}
\lref\gnttown{J. P. Gauntlett, G. W. Gibbons, G. Papadopoulos and P.K.
Townsend, {\it ``Hyper-K\"ahler Manifolds and Multiply Intersecting 
Branes''}, Nucl. Phys. {\bf B500} (1997) 133; hep-th/9702202.} 
\lref\witfourd{E. Witten, {\it ``Solutions of Four Dimensional Field 
Theories Via M-Theory''},  Nucl. Phys. {\bf B500} (1997) 3;
hep-th/9703166.} 
\lref\givkut{A. Giveon and D. Kutasov, {\it ``Brane Dynamics and Gauge
Theories''}; hep-th/9802067.} 
\lref\hanzaf{A. Hanany and A. Zaffaroni, {\it ``On The Realization of
Chiral Four Dimensional Gauge Theories using Branes''}, JHEP {\bf 05}
(1998) 001; hep-th/9801134.}
\lref\hanuranga{A. Hanany and A. Uranga, {\it ``Brane Boxes and Branes 
on Singularities''}, JHEP {\bf 05} (1998) 013; hep-th/9805139.}
\lref\ksil{S. Kachru and E. Silverstein, {\it ``4D Conformal Field
Theories and Strings on Orbifolds''}, Phys. Rev. Lett. {\bf 80}
(1998) 4855; hep-th/9802183.} 
\lref\lnv{A. Lawrence, N. Nekrasov and C. Vafa, {\it ``On Conformal
Field Theories in Four Dimensions''}, Nucl. Phys. {\bf B533} (1998)
199; hep-th/9803015.}
\lref\dfk{U. Danielsson, G. Ferretti and I. Klebanov, {\it ``Creation
of Fundamental Strings from Crossing D-Branes''},
Phys. Rev. Lett. {\bf 79} (1997) 1984; hep-th/9705084.} 
\lref\dmconif{K. Dasgupta and S. Mukhi, {\it ``Brane Constructions,
Conifolds and M-Theory''}, hep-th/9811139, Nucl. Phys. B, to appear.}
\lref\calmal{C. Callan and J. Maldacena, {\it ``Brane Dynamics from 
Born-Infeld Action''}, Nucl. Phys. {\bf B513} (1998) 198; hepth/9705084.}
\lref\sennonbps{A. Sen, {\it ``Stable Non-BPS States in String Theory''},
JHEP {\bf 06} (1998) 007; hep-th/9803194\semi
A. Sen, {\it ``Stable Non-BPS Bound 
States of BPS D-Branes''}, JHEP {\bf 08} (1998) 010, hep-th/9805019\semi
A. Sen, {\it ``Tachyon Condensation on Brane-Antibrane Systems''},
JHEP {\bf 08} (1998) 012, hep-th/9805170.}    
\lref\wittenk{E. Witten, {\it ``D-Branes and K-Theory''},
JHEP {\bf 12} (1998) 019; hep-th/9810188. }
\lref\othernonbps{E. Gava, K.S. Narain and M.H. Sarmadi, {\it ``On the 
Bound States of $p$-Branes and $p+2$-Branes''},
Nucl. Phys. {\bf B504} (1997) 214; hep-th/9704006\semi
O. Bergman and M. Gaberdiel, {\it ``Stable Non-BPS D-Particles''},
Phys. Lett. {\bf B441} (1998) 133; hep-th/9806155\semi
P. Horava, {\it ``Type IIA D-Branes, K-Theory and Matrix Theory ''};
hep-th/9812135.} 
\lref\sencyc{A. Sen, {\it ``BPS D-Branes on Non-supersymmetric
Cycles''}; hep-th/9812031.}
\lref\bergcyc{O. Bergman and M. Gaberdiel, {\it ``Non-BPS States in
Heterotic-Type IIA Duality''}; hep-th/9901014.}
\lref\aspinwall{P. Aspinwall, {\it ``Enhanced Gauge Symmetries and K3
Surfaces''},  Phys. Lett. {\bf B357} (1995) 329; hep-th/9507012.}
\lref\bdm{C. P. Bachas, M. Douglas, M. B. Green, {\it ``Anomalous 
Creation of Branes''}, JHEP {\bf 07} (1997) 002; hep-th/9705074.}
\lref\djm{K. Dasgupta, D. Jatkar and S. Mukhi, {\it ``Gravitational
Couplings on $Z_2$ Orientifolds''}, Nucl. Phys. {\bf B523} (1998) 465; 
hep-th/9707224.}
\lref\bbg{C. Bachas, P. Bain and M.B. Green, {\it ``Curvature Terms in
D-Brane Actions and their M-Theory Origin''}; hep-th/9903210.}
\lref\romans{L. J. Romans, {\it ``New Compactifications of Chiral
$N=2$, $d=10$ Supergravity''}, Phys. Lett. {\bf 153B} (1985) 392.}
\lref\vonunge{R. Von Unge, {\it `` Branes at Generalised Conifolds 
and Toric Geometry''}, JHEP {\bf 02} (1999) 023; hep-th/9901091.}
\lref\lopez{E. Lopez, {\it ``A Family of $\cN=1$ $SU(N)^k$ Theories from
Branes at Singularities''}, JHEP {\bf 02} (1999) 019; hep-th/9812025.} 
\lref\piljin{Piljin Yi, {\it ``Membranes from Five-Branes and Fundamental
Strings from Dp-Branes''}; hep-th/9901159.}
\lref\afhs{B.S. Acharya, J.M. Figueroa-O'Farrill, C.M. Hull and
B. Spence, {\it ``Branes at Conical Singularities and Holography''},
hep-th/9808014.}
\lref\karch{M. Aganagic, A. Karch, D. Lust, A. Miemiec, {\it ``Mirror 
Symmetry for Brane Configurations and Branes at Singularities''}; 
hep-th/9903093.}

{\nopagenumbers
\Title{\vtop{\hbox{hep-th/9904131}
\hbox{IASSNS-HEP-99/38}
\hbox{TIFR/TH/99-16}}}
{\vtop{\centerline{Brane Constructions, Fractional Branes}
\centerline{}
\centerline{and Anti-deSitter Domain Walls}}}
\centerline{Keshav Dasgupta\foot{E-mail: keshav@ias.edu}}
\vskip 3pt
\centerline{\it School of Natural Sciences, Institute for Advanced Study}
\centerline{\it Olden Lane, Princeton NJ 08540, U.S.A.}
\vskip 7pt
\centerline{Sunil Mukhi\foot{E-mail: mukhi@tifr.res.in}}
\vskip 3pt
\centerline{\it Tata Institute of Fundamental Research,}
\centerline{\it Homi Bhabha Rd, Mumbai 400 005, India}

\ \smallskip
\centerline{ABSTRACT}

Compactifications of type IIB string theory on $AdS_5 \otimes X_5$,
where $X_5$ is an Einstein space, can have one-fourth or half maximal
supersymmetry for certain choices of $X_5$. Some of these theories
admit exotic domain walls arising from 5-branes wrapping 2-cycles in
$X_5$. We explore the relationship among these domain walls,
fractional branes and branes stretched on intervals. World-volume
fluxes in the wrapped branes play an important role in the analysis.
We draw some parallels between the $AdS$ background with exotic domain
walls and $\cN=1$ supersymmetric MQCD, and identify other extended
objects on the $AdS$ side in the dual brane construction. The
process of brane creation is used to give an alternate derivation of
the relationship between fractional branes and branes on intervals.

\Date{April 1999}
\vfill\eject}
\ftno=0

\newsec{Introduction}

The study of branes at singularities has provided remarkable new
insight into geometry and gauge theory. Within this circle of ideas,
one should include the study of field theories on branes, both at
quotient singularities\refs{\dougmoore,\dgm}\ and at
non-quotient singularities such as conifolds\refs\klebwit, and the
large-N limits of these two classes of configurations which give rise
to string and M-theory compactifications on $AdS_p \otimes
X_q$\refs{\malda,\ksil,\lnv}. Recently, useful knowledge about these
models has been obtained using brane constructions, following the
original idea of Hanany and Witten to realise field theories by
suspending branes between branes.

The simplest example of the above class of models is to consider a
smooth space transverse to the brane, (a trivial singularity, so to
speak). For D3-branes in type IIB string theory, the transverse space
is $R^6$ and the large-N limit is believed to be dual to the string
theory on $AdS_5 \otimes S^5$. This arises because $R^6$ is a half-line
fibred over an $S^5$ whose size varies with distance from the origin,
and the large-N limit ``blows up'' the varying $S^5$ to instead have a
constant size. This leads naturally to the expectation that the large-N
limit will have a nontrivial effect on any singularity that one may
introduce. 

Continuing to work with D3-branes in type IIB string theory, the
simplest singularity is a $Z_n$ ALE space. Here the transverse space
to the branes is $R^2\otimes (R^4/Z_n)$ where $Z_n$ has the obvious
action as a subgroup of $SU(2)$ on the complex coordinates of
$R^4$. In this case, there is an entire fixed plane of singularities
transverse to the branes. In the large-N limit, this singular locus
gets reduced to the intersection of the fixed plane and an $S^5$,
which is a fixed circle. The spacetime description becomes $AdS_5
\otimes (S^5/Z_n)$. Half the supersymmetries are broken in this case,
as compared with the previous case, thus the D3-brane gauge theory has
$N=2$ supersymmetry. 

This example can be extended to $\cN=1$ supersymmetry in two distinct
ways. One is to quotient $R^6$ by the action of a group $\Gamma$ that
sits naturally in $SU(3)$, for example $Z_3$ or $Z_k\otimes
Z_{k'}$. For the $Z_3$ case, the transverse space to the branes
initially has a point singularity at the origin, hence the large-N
limit gives rise to compactification on a completely smooth space. In
the latter case, there are typically intersecting singularities both
before and after taking the large-N limit, though in special cases
the space may be smooth.

The other way to extend these examples is to consider D3-branes
transverse to more general singularities that are not quotient
singularities. The prototype of these examples is the conifold
singularity of a Calabi-Yau manifold. In this case one again gets
$\cN=1$ supersymmetry on the brane world-volume. The transverse space
has a point singularity, the node of the conifold, while in the
large-N limit one has string theory on $AdS_5 \otimes T_{1,1}$ where
$T_{1,1}$ is a smooth Einstein space\refs\romans\ which is the
``base'' of the conifold\refs\candelas. The CFT on the D3-brane at a
conifold was analysed in Ref.\refs\klebwit\ and, via brane
constructions, in Refs.\refs{\angel,\dmconif}. 

There are more general 6-dimensional singularities than the conifold
(the A-D-E generalized conifolds were first described in
Ref.\refs\gjm\ in the context of noncritical strings, and subsequently
studied in the brane context in Refs.\refs{\bsv,\oogvaf}), which tend
to have extended singular loci rather than a single node. The large-N
limits of these cases have also been
investigated\refs{\afhs,\mori,\angel,\gns,\lopez,\vonunge}.

There are other quotients of the basic configuration which correspond
to orientifolding. Though we will not discuss them here, they too,
present many interesting features.

In all the examples described above, one can give a dual ``brane
construction'' by performing a suitable T-duality. For example, the
theory of N D3-branes at a $Z_n$ ALE space is T-dual to a
configuration of $n$ parallel NS 5-branes in type IIA string theory,
with N D4-branes stretching between them\refs{\acl,\kls}. This is the
so-called $\cN=2$ ``elliptic'' model\refs\witfourd. For more general
quotient singularities one gets ``brane box''
configurations\refs{\hanzaf,\hanuranga}. In these models the spectrum
and the superpotential can be read out using definite rules given in
Ref.\refs{\hanzaf}.
(However, as was shown 
recently in an interesting paper by Aganagic et. al.\refs{\karch},
these rules, according to which the superpotential is obtained 
by drawing closed
triangles across the boxes, are not true in general. To get the right
quartic superpotential one has to use the ``diamond'' 
rule\refs{\karch}\foot{This resolves a puzzle that was noted 
in Ref.\refs{\dmconif}.}.
These rules reduce to the triangle rules only in special
circumstances.)
Similarly, the theory of N
D3-branes at a conifold is dual to a configuration of perpendicular NS
5-branes with N D4-branes stretching between
them\refs{\angel,\dmconif}. Hence it is possible to look for direct
correspondences between brane constructions and $AdS$
compactifications.

Since we will be interested in relating the above ideas to domain
walls, let us briefly review the relevant known results. For
concreteness, consider first the system of N D3-branes at a
conifold. In the $AdS_5 \otimes T_{1,1}$ model, it was
argued\refs\gubkleb\ that there are two kinds of domain walls in the
$AdS_5$ spacetime. A domain wall in 5-dimensional spacetime must be
some kind of 3-brane. In the present case one can introduce the
D3-branes of type IIB string theory, or one can take a D5-brane and
wrap it on a 2-cycle of the Einstein space $T_{1,1}$.  We will refer to
the latter kind of domain wall as ``exotic''. (Alternatively one can
also wrap an NS 5-brane, which is dual to the D5-brane under
S-duality.) Domain walls can be oriented in different ways within
$AdS$, we will consider the orientations which preserve all the
supersymmetry (thus the brane is parallel to the $AdS_5$ boundary).

The key property of domain walls in $AdS$ compactifications is that
the radius of the $AdS$ space jumps when one crosses them. In terms of
the dual gauge theory, the rank of the gauge group increases or
decreases across a domain wall. For the maximally supersymmetric
$AdS_5 \otimes S^5$ compactification, the gauge group goes from
$SU(N)$ to $SU(N+1)$ across a domain wall made up of a D3-brane. This
is the only possible domain wall in this case, since $S^5$ has no
2-cycles.

In the $AdS_5\otimes T_{1,1}$ case, the gauge group of the CFT is
$SU(N)\otimes SU(N)$. Inserting a D3-brane changes the gauge group to
$SU(N+1)\otimes SU(N+1)$. However, the exotic domain wall obtained by
wrapping a D5-brane on a 2-cycle of $T_{1,1}$ ($T_{1,1}$ is
topologically the same as $S^2\times S^3$) changes the gauge group to
$SU(N+1)\otimes SU(N)$.

The standard domain wall is easy to understand in the brane
construction: its effect is to add an extra D4-brane in the elliptic
models, wrapping all the way around the compact direction. As we will
argue in some detail, the exotic domain wall corresponds to a D4-brane
stretching part of the way around a compact direction and then ending
on parallel NS 5-branes (this has been independently noted in
Ref.\refs\gns).

That such an object can correspond to a wrapped D5-brane in the T-dual
picture is initially surprising, since the T-duality direction would
appear to convert the D4 to a D3-brane. However, this discrepancy of 2
dimensions is resolved by the observation that such branes in a brane
construction are actually ``fractional branes''\refs\fractional\ of
the T-dual theory. Fractional branes are interpreted as wrapped
branes of 2 higher dimensions, wrapping a 2-cycle of vanishing size
that is hidden in the singularity. They are supposed to acquire their
charge and tension from a flux of the 2-form field $\bns$ through the
vanishing 2-cycle, though we will see below that this is not the
complete explanation.

Since the large-N limit for N D3-branes at a conifold effectively
blows up a vanishing 2-cycle of the conifold to a finite $S^2$, the
fractional branes at a conifold blow up into exotic domain
walls. Since fractional branes have already been associated in
Refs.\refs{\kls}\ with branes stretched on an interval, the relation
between the latter objects and domain walls seems quite
natural. However, we will see that world-volume field strengths on the
wrapped branes play a crucial role in providing the right properties
that are predicted by various dualities.

We will also provide an independent argument for the relationship
between domain walls and stretched branes using the process of ``brane 
creation''\refs\hanwit. 

This paper is organized as follows. In Section 2 we review the
properties of 3-branes at a $Z_2$ ALE space and explain how fractional
branes arise and how they are described in a T-dual picture. Although
much of this material is known, we clarify the role of world-volume
fluxes and the relationship of this system to the process of tachyon
condensation via a vortex solution on a brane anti-brane pair, which
has been extensively discussed of
late\refs{\sennonbps,\wittenk,\othernonbps}. The tachyon is actually
absent in the limit relevant to our problem, but the field strength
associated to the would-be vortex causes the pair to annihilate into a
lower brane. In Section 3 we extend these considerations to the
conifold singularity. In particular, we observe that the K\"ahler
transition at $\bns=0$, T-dual to the crossing of two NS 5-branes,
corresponds to a jump in the total world-volume field strength. In
Section 4 we turn to the large-N limit and the associated exotic
domain walls. These are identified with a T-dual description of branes
on an interval. We provide a more complete picture of the jump in the
gauge group across such domain walls in the light of our observations
about world-volume fluxes. We also relate the model in the presence of
such domain walls to MQCD. In Section 5 we examine other wrapped
branes in the $AdS_5 \otimes T_{1,1}$ model, and describe the
corresponding objects in the T-dual brane construction.  Finally, in
Section 6 we provide an alternate derivation of the relationship
between domain walls and branes on intervals using the phenomenon of
brane creation.

\newsec{D4-Branes on an Interval: the $\cN=2$ Case}

Consider a pair of NS 5-branes in type IIA theory. Let us start by
taking them parallel, filling the directions $(x^1,\ldots,x^5)$,
coincident in $(x^7,x^8,x^9)$ and separated by a finite interval along
$x^6$. Take the $x^6$ direction to be compact. Now take a D4-brane
that terminates on each of the NS 5-branes as a 3-brane along
$(x^1,x^2,x^3)$, and stretches between them along $x^6$ (Fig. (2.1)).
Next we T-dualize this circle.
\bigskip

\centerline{\epsfbox{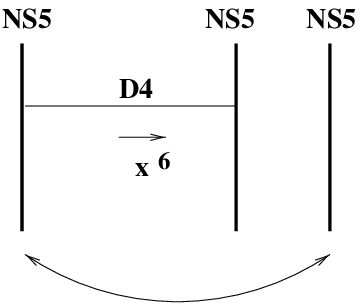}}\nobreak
\centerline{Fig. (2.1): D4-Brane on an Interval}
\bigskip

The result of this T-duality was considered in Ref.\refs\kls\ which
used the fact that NS 5-branes turn into Kaluza-Klein monopoles
of type IIB theory. Coincident KK monopoles describe the near-horizon
behaviour of an ALE singularity.  As they are T-dual to type IIA
NS 5-branes, they carry tensor multiplets on their world-volumes. In
\refs\kls\ it was proposed that the D4-brane on an interval turns
into a fractional brane\refs\fractional\ at the ALE singularity.

To see this, let us start with a singular $Z_2$ ALE space\foot{The
considerations in this section have a straightforward extension to the
case of $Z_n$, although extending them to the other discrete groups of
$D_n$ or $E_n$ type could be somewhat nontrivial.}\ along directions
$x^6,x^7,x^8,x^9$. The node is really a 5-plane filling the remaining
directions. Close to the singular point, the space can be replaced by
a 2-centre Taub-NUT metric with coincident centres. This
is equivalent to saying that we have two coincident Kaluza-Klein
monopoles. We also know\refs\aspinwall\ that the $Z_2$ orbifold hides
half a unit of $B_{NS,NS}$ flux through the shrunk 2-cycle
$\Sigma$. The four moduli associated to this ALE space are three
geometrical parameters, which can be thought of as the blowup of the
ALE to form a smooth Eguchi-Hanson metric, and the $B_{NS,NS}$ flux.

Take a D3-brane transverse to the ALE space, filling the
directions $x^1,x^2,x^3$. (More generally we start with N such
D3-branes.) When the ALE space is singular, the
world-volume theory of the 3-brane has two branches: a Higgs branch,
when the brane is separated from the singularity along 
$x^6,x^7,x^8,x^9$, and a Coulomb branch when the brane hits the
singularity and dissociates into a pair of ``fractional branes'' which
can move around only in the $x^4,x^5$ directions. However, if the ALE
space is blown up, then the Coulomb branch is lifted.

The fractional branes are interpreted as a pair of D5-branes whose
5-brane charges cancel (hence they are really a D5-brane --
anti-D5-brane pair). However, they carry 3-brane charge by virtue of
the CS coupling on D5-branes. Denoting the world-volume gauge field
strength on the D5-brane by $F_1$, we have the coupling
\eqn\cscoupling{
\int (\bns - F_1)\wedge D^+ }
where $D^+$ is the self-dual 4-form potential in the type IIB
string. At the orbifold point we have $\int_\Sigma \bns = \half$ and
hence half a unit of D3-brane charge. There is an apparent puzzle here,
since the anti-5-brane (whose world-volume gauge field strength is
$F_2$) will have a coupling
\eqn\antics{
-\int (\bns - F_2)\wedge D^+ }
Hence we would expect the anti-5-brane to acquire a $-\half$ unit of
D3-brane charge in this way, resulting in a net D3-brane charge of 0,
while we require it to be 1. The resolution of this puzzle comes from
the fact that the B-field flux is not gauge-invariant. By gauge
transformations, it can be effectively made into a periodic variable
taking values from 0 to 1. The correct gauge-invariant quantity on a
brane is $\bns - F$ which appears in the couplings above.

Hence what really happens is the following. Viewed as a spacetime
field, $\bns$ has a flux of $\half$ (which is the same as $-\half$
since it is a periodic variable). {}From Eqs.\cscoupling,\antics, this
contributes $\half$ a unit of D3-brane charge to the wrapped 5-brane
and $-\half$ to the anti-5-brane. Now let us also turn on a
world-volume gauge field strength $F_2$ on the anti-brane and give it
a flux of $+1$ unit through the vanishing 2-cycle $\Sigma$ (more
generally, we assign unit flux to the relative gauge field $F_- = F_2
- F_1$). In this configuration, the 5-brane anti-5-brane pair has
total 3-brane charge equal to 1.

Examining Eqs.\cscoupling\ and \antics\ above, we see that variations
of the B-field act equally and oppositely on the brane and
anti-brane. This changes the relative 3-brane charge on each of them
in such a way that the sum remains constant. Thus it is not quite
correct to say that the total 3-brane charge and tension of the brane
anti-brane pair arise from the B-flux. Such a statement is valid only
if restricted to an individual brane, where one can always gauge away
the world-volume field strength by a gauge transformation on $\bns$.

Next, perform a T-duality along $x^6$. Then the pair of Kaluza-Klein
monopoles turns into a pair of NS 5-branes in type IIA string
theory. We expect that the $B$-field turns into a geometrical
modulus. In fact, it has been argued \refs\kls\ that it becomes the
separation of the NS 5-branes along the $x^6$ direction. The way to see
this is that a D3-brane at a $Z_2$ ALE space has the following
terms\refs\dougmoore\ in its world-volume action\foot{It is
known\refs\djm\ that couplings of this type on D3-branes receive
instanton corrections which convert them into modular forms under
SL(2,Z) S-duality (a recent discussion can be found in
Ref.\refs\bbg). It would be interesting to understand the analogous
corrections in the present case.} which depend on the fluxes of the
2-form fields $\bns$ and $\brr$ through the vanishing cycle of the ALE
space:
\eqn\follterms{
\int \half(b_1\, F \wedge *F + b_2\, F\wedge F)}
where
\eqn\beedef{
b_1 \sim \int_\Sigma \bns,\qquad
b_2 \sim \int_\Sigma \brr}
The above formula is actually valid for one fractional 3-brane (say
the wrapped D5-brane) and we should replace $\bns$ by $\bns - F_1$. An
analogous formula holds for the other fractional 3-brane (the wrapped
anti-D5-brane) with $\bns - F_1$ replaced by $-(\bns - F_2)$.  It
follows that, with $\int_\Sigma F_1=0$ and $\int_\Sigma F_2=1$, the
gauge couplings of the two $U(1)$ gauge groups are given by
$\int_\Sigma \bns$ and $(1-\int_\Sigma \bns)$ respectively. Hence
T-dualizing along this direction produces a pair of NS 5-branes in
type IIA theory whose separation along the $x^6$ circle is
proportional to $\int_\Sigma \bns$ (in the other direction the
separation is therefore proportional to $(1-\int_\Sigma \bns)$).

The mapping between parameters has been discussed in Ref.\refs\kls. We
will review and extend this analysis in the light of our observations
about the role of world-volume gauge field fluxes on the brane.  In
the brane construction, the two NS 5-branes can move around on the
$x^6$ circle or they can separate from each other along
$x^7,x^8,x^9$. These four possible motions must correspond, on the
orbifold side, to the four deformation parameters associated to the
orbifold string theory: three geometrical deformations (two complex
and one K\"ahler) and a $\bns$-field modulus.  On the Higgs branch,
the D3-brane is separated from the ALE singularity along
$x^6,x^7,x^8,x^9$. In the T-dual type IIA picture, the 4-brane is
separated from the NS 5-branes along $x^7,x^8,x^9$ and has a Wilson
line along $x^6$ in its world-volume. In particular, even taking
$x^7=x^8=x^9=0$, so that the D4-brane touches the NS 5-branes, it
cannot split into pieces stretching on intervals as long as there is a
Wilson line.

Going to the Coulomb branch, by tuning all of $x^6,x^7,x^8,x^9$ to 0,
the picture is somewhat different. At this point the D3-brane splits
into a pair of fractional branes which can move independently along
$x^4,x^5$. The geometric orbifold singularity now cannot be blown up
any more. This is easy to see on the T-dual type IIA side, where the
D4-brane splits into two pieces that stretch along the two intervals
between the two NS 5-branes (one from each side of the $x^6$ circle).
These partially wrapped 4-branes can move independently along the
NS 5-branes, namely in the $x^4,x^5$ directions (Fig. (2.2)). 
\bigskip

\centerline{\epsfbox{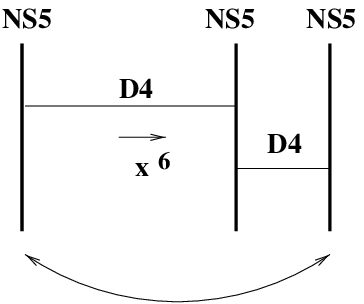}}\nobreak
\centerline{Fig. (2.2): Wrapped D4-Brane Splits in Two}
\bigskip

When the D4-brane segments are separated, the NS 5-branes cannot be
separated along $x^7,x^8,x^9$ without some cost in energy. This is
dual to the statement in the type IIB theory that once a D3-brane has
reached the orbifold and split into fractional branes (i.e. on the
Coulomb branch), the orbifold singularity can no longer be resolved
while preserving supersymmetry.

We note in passing that since the D4-brane can split into two pieces
which can separate to arbitrary distances, by cluster decomposition we
would expect each separate piece (a D4-brane stretched on an interval)
to be a valid state in the theory\refs\fractional. In type IIB
language, this is equivalent to saying that a single fractional brane
makes sense, though the process of bringing a D-brane to a $Z_2$
singularity always generates fractional branes in pairs.

On the Coulomb branch, the NS 5-branes are still free to move around
in the $x^6$ direction, as they were in the Higgs branch, with the
position being T-dual to the flux of $\bns$. Now we see that this
motion also changes the relative tensions of the D4-branes stretched
along intervals (we are referring to the tensions from a
(3+1)-dimensional viewpoint), though the sum remains constant. These
two facts are mutually consistent only because, as we have pointed out
above, the B-field couples oppositely to the two fractional branes as
exhibited in Eqs.\cscoupling,\antics\foot{Thus in particular, we
disagree with a claim made in Ref.\refs\kls\ (below Eq.(12)) that the
motion of the NS5-branes in the type IIA picture is related to Wilson
lines.}.

To summarise, the type IIB picture on the Coulomb branch is that the
relative world-volume gauge field strength $F_-$ on the 5-brane
anti-5-brane pair must be turned on over the 2-cycle $\Sigma$ and
gives rise to a 3-brane in the space transverse to that cycle. The
spacetime $\bns$ flux over $\Sigma$ changes the relative tensions of
the wrapped brane and anti-brane keeping the total constant. In the
T-dual type IIA picture, the $\bns$ flux corresponds to the relative
location of the NS 5-branes along $x^6$, which has the same effect on
the 3-brane tensions of the two finitely extended D4-brane segments.

This phenomenon is reminiscent of tachyon condensation, discussed in
Refs.\refs{\sennonbps,\wittenk,\othernonbps}. It is interesting to
compare the two. The phenomenon in Refs.\refs\sennonbps\ involves a
$p$-brane anti-$p$-brane pair that supports a tachyon. One then allows
the tachyon to condense in a nontrivial configuration, in other words
to develop a VEV which depends on only two spatial directions and
resembles a vortex solution. The tachyon is charged under the relative
gauge field $F_-$ and its condensation is accompanied by an excitation
of $F_-$ with unit flux through a 2-cycle. (This 2-cycle is usually
taken to be infinite 2-space or else a 2-torus.)  The condensation
phenomenon takes us from an unstable configuration consisting of a
$p$-brane anti-$p$-brane pair, to a stable configuration of a
unit-charged $(p-2)$-brane. Multi-vortices create multi-charged
$(p-2)$-branes.

Our phenomenon involving fractional branes also gives rise to a unit
charged $(p-2)$-brane from a $p$-anti-$p$ pair (in this case
$p=5$). Again, the lower brane charge is created by a flux of $F_-$ in
the world-volume theory. However, in this case the 2-cycle over which
$F_-$ acquires a flux is of zero size. Related to this is the fact
that there is no tachyon in this limit\foot{An argument for this (due
to Ashoke Sen) is the following: Consider a brane anti-brane pair
wrapped on a 2-torus, with a 0-brane charge of 2 units. Under
T-duality, this turns into a pair of 2-branes wrapped on the dual
torus, with a nontrivial $SU(2)$ field strength through it. The fact
that this is not a configuration of minimum energy in its sector
corresponds to the tachyonic instability. However, as the original
torus shrinks, the dual torus becomes infinite and the $SU(2)$ field
strength (since it has constant flux through the dual torus) goes to
zero. Thus in the limit, there is no tachyon.}. Hence, going from the
brane anti-brane pair to a lower-dimensional brane does not lower
energy, but instead corresponds to a marginal deformation. This means,
for example, that one can also go in the opposite direction: a
D3-brane in the plane of the ALE space can split into a D5 anti-D5
pair, which is precisely the phenomenon we started out to
discuss. Additionally, the {\it distribution} of 3-brane charge and
tension between 5-brane and anti-5-brane can be varied by
turning on the spacetime B-field, which keeps the {\it total} 
3-brane charge fixed.

Thus, in the $\cN=2$ supersymmetric example discussed above, of
D3-branes at an ALE space, we see how a partially wrapped brane in
type IIA theory, becomes a fractional brane in type IIB. Though this
has already been discussed, for example, in Ref.\refs\kls, we have
identified more precisely the interplay between the world-volume gauge
fields and the $\bns$ flux, which is essential to obtain a consistent
picture. World-volume fluxes on the brane anti-brane pair at an
orbifold singularity were used in Refs.\refs{\sencyc,\bergcyc}\ in
configurations where the $\bns$ flux was fixed at the ``orbifold''
value of $\half$. Here, motivated by the brane construction dual, we
have explained how the $\bns$ flux really determines only the relative 
tensions of the pair. Some consequences of this fact will be relevant
in Section 4.

Our final observation about this system has to do with a process in
which the two NS5-branes connected by D4-branes pass through each
other\foot{This observation emerged in discussions with Ashoke
Sen.}. Recall the case discussed above, of a D4-brane going around
$x^6$, which has separated into two segments (Fig. (2.2)). Let
these be well-separated from each other. Now let us bring the two
NS5-branes together. In this limit, one of the segments wraps the
entire circle while the other shrinks to zero length (Fig. (2.3)).
\bigskip

\centerline{\epsfbox{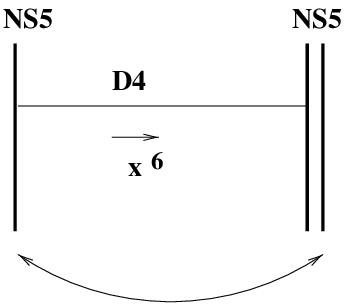}}\nobreak
\centerline{Fig. (2.3): Two NS5-Branes Coinciding}
\bigskip

However, if we continue past this point and let the NS5-branes cross,
then the first D4-brane segment wraps once around the circle and then a
part of the way again. Meanwhile the other segment, which had shrunk
to zero size, grows back with the {\it opposite} orientation
(Fig. (2.4)).
\bigskip

\centerline{\epsfbox{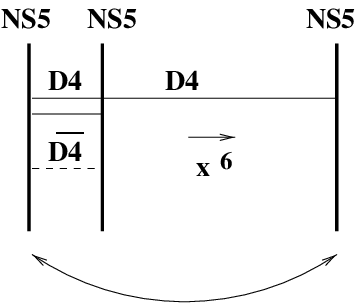}}\nobreak
\centerline{Fig. (2.4): Two NS5-Branes Cross (the dotted line is an 
anti-D4-brane)}
\bigskip

The result is that along one direction between the NS5-branes, there
is a pair of segments of a D4-brane and an anti-D4-brane. These can
annihilate, so the configuration is unstable and supersymmetry must be
broken.

In the dual picture of fractional 3-branes, one of them goes to zero
tension and then comes back as an anti-3-brane, while the other is like
a marginally bound state of a D3-brane and a fractional 3-brane.
Clearly this is a non-supersymmetric configuration.

\newsec{D4-Branes on an Interval: the $\cN=1$ Case}

Let us now generalize this to an $\cN=1$ example: D3-branes at a
conifold singularity. This singularity is localized at a point (the
origin) in the $(x^4,\cdots,x^9)$ directions. The analysis is quite
similar to the previous case, except that there is really no Coulomb
branch except the origin. The T-dual picture of the conifold is a pair
of NS 5-branes of type IIA string theory, as for the $Z_2$ ALE space,
except that one of the 5-branes is rotated with respect to the
other\refs{\angel,\dmconif}. Thus the first NS 5-brane fills the
directions $(x^1,x^2,x^3,x^4,x^5)$ as in the ALE case, while the
second one, conventionally called an NS' 5-brane, stretches along
$(x^1,x^2,x^3,x^8,x^9)$. As before, they are separated along $x^6$
which is the T-duality direction (Fig (3.1)).
\bigskip

\centerline{\epsfbox{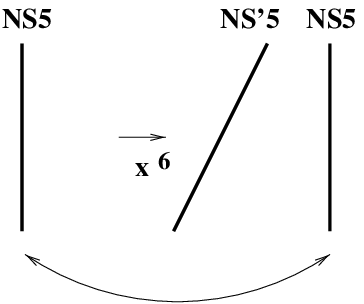}}\nobreak
\centerline{Fig. (3.1): Brane Construction Dual to Conifold}
\bigskip

This time the process of bringing a D3-brane (which fills the
$(x^1,x^2,x^3)$ directions) to the conifold involves tuning the 6
coordinates $(x^4,\cdots,x^9)$ to zero. In the T-dual type IIA
description, the process consists of taking a D4-brane that fills
$(x^1, x^2, x^3, x^6)$, and bringing it to a fixed set of values of
the remaining 5 coordinates -- these are the $(x^8,x^9)$ locations of
the NS 5-brane, the $(x^4,x^5)$ locations of the NS' 5-brane and the
common $x^7$ location of both of them. Finally, a Wilson line in the
D4-brane world-volume replaces the $x^6$ location of the D3-brane on
the type IIB side.

Now, unlike in the $\cN=2$ case, the D3-brane cannot really split into
fractional branes which move apart. The reason is clear in the dual
picture: the rotation on one NS 5-brane has lifted the Coulomb
branch. Nevertheless, it is consistent to think of the D4-brane which
wraps around the $x^6$ direction as being made up of a pair of
fractional branes which stretch between the NS 5-brane and the NS'
5-brane along opposite sides of the $x^6$ circle. Locally, near one
of the NS 5-branes, the D4-brane will behave exactly as for the $\cN=2$
supersymmetric case discussed in the previous section (this fact was
exploited in Ref.\refs\dmconif\ to extract the spectrum and
symmetries of the $\cN=1$ theory). Moreover, at the level of
world-volume gauge theory too, one can see the origin of the Coulomb
branch: the gauge group has two $SU(N)$ factors (assuming we brought
N D3-branes to the conifold).

The main difference between this case and the ALE singularity is that
here the D4-brane charge does not need to ``flow'' onto the NS 5-brane,
but passes right through and travels around the $x^6$ circle. In the
M-theory limit, this means that the D4 and NS 5-branes do not unify
into a single M5-brane, but remain three separate
components\refs\dmconif.

It is clear from the geometry in the type IIA T-dual picture that one
can vary the gauge couplings by moving the NS 5-branes along the
compact $x^6$ direction. Just as for the ALE case, this corresponds to
changing the relative sizes of two fractional branes, even if here
this is not very meaningful because the original D3-brane cannot
really split into separated fractional branes. Nevertheless the
phenomenon is one that we have encountered earlier, that of turning on
a unit flux of $F_-$ in the world-volume theory of a D5 anti-D5 pair,
and then varying the spacetime $\bns$ flux. These fluxes are now
through the 2-cycle of the conifold that has shrunk to zero size.

Because of the absence of a Coulomb branch, the interpretation in
terms of fractional branes may appear somewhat trivial in the conifold
case. However, it is possible for us to {\it add} a single fractional
brane to this picture. On the type IIB side, we consider the state in
which a single D5-brane wraps the vanishing 2-cycle of the
conifold. Just as in the case of the ALE singularity, there is a
special point in moduli space where this 2-cycle naturally conceals
half a unit of $\bns$ flux, so the result is half a 3-brane. On the
type IIA side, this is one extra D4-brane stretching between the NS
5-brane and NS' 5-brane along one side of the $x^6$ circle only
(Fig. (3.2)).
\bigskip

\centerline{\epsfbox{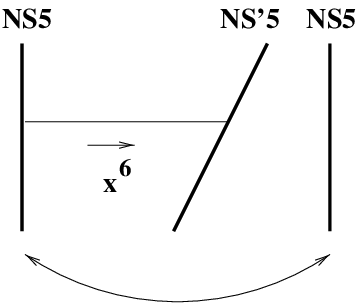}}\nobreak
\centerline{Fig. (3.2): Fractional Brane at a Conifold}
\bigskip

 Thus we see that the identification between fractional
branes and branes on an interval holds for the conifold theory too.

This has an interesting consequence: we had noted above that in the
M-theory limit, a D4-brane wrapped along $x^6$ becomes an M5-brane
which passes through the orthogonal M5-branes at fixed locations in
$x^6$. With the addition of a fractional brane, this no longer
happens. The fractional brane will correspond, in the M-theory limit,
to an M5-brane that ``flares out'' and joins smoothly onto the
orthogonal M5-branes (Fig. (3.3)). As a result, the M-theory
configuration which in the absence of a fractional brane consisted of
three disconnected sets of M5-branes, becomes joined in an obvious way
as soon as a single fractional brane is added to the system. The
M5-branes which wrap around $x^6,x^{10}$ are unaffected but the two
M5-branes which in the type IIA limit correspond to the NS5 and NS5'
branes respectively, get linked to each other. This reduces the
symmetries of the (3+1)-d field theory and introduces dynamical
effects into the model similar to those studied in the context of
$\cN=1$ supersymmetric QCD in Ref.\refs\witqcd. We will examine some
of these effects in subsequent sections.
\bigskip

\centerline{\epsfbox{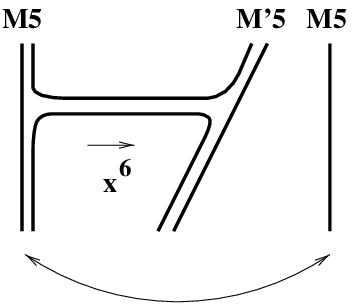}}
\centerline{
\vtop{
\hbox{Fig. (3.3): M-Theory Limit of Fractional}
\hbox{\phantom{Fig. (3.3): }Brane at a Conifold}}}
\bigskip

The issues discussed above also have a bearing on the ``K\"ahler
transition'' discussed in the context of the brane construction dual
to generalized conifolds, in Ref.\refs{\angel,\vonunge}. This is a
``phase transition'' in a limited sense: by separating the branes
along the $x^7$ direction one can go around this point in moduli
space, but if we keep the $x^7$ locations equal then we pass through a
singularity. In Ref.\refs\angel\ it was pointed out that on the
conifold (type IIB) side, this transition is a non-geometrical
analogue of the well-known ``flop'' transition (in the latter case we
would keep the 5-branes coincident along $x^6$ and change their $x^7$
separation until they pass through each other). Unlike the flop,
obtained by formally varying a $P^1$ from positive to ``negative''
size, the present K\"ahler transition arises by varying the $\bns$
flux from a positive to a negative value.

{}From our discussion, we get some new insight into this process.  We
will confine ourselves to the simple conifold and ask what happens
when the NS5-brane passes through the NS' 5-brane. In the absence of
D4-branes stretching between the 5-branes, this transition is trivial:
the configuration of an NS5-brane followed by an NS' 5-brane is
equivalent to the one with reverse ordering by an overall translation
along $x^6$. However, when a D4-brane is stretched around the $x^6$
circle, something more nontrivial happens. As we have seen, this
D4-brane is a wrapped D5 anti-D5 pair in the type IIB
description. Suppose initially $\int_\Sigma \bns=\epsilon$ where
$\epsilon$ is a small positive number. Now vary $\int_\Sigma \bns$ so
that it passes through zero and becomes $-\epsilon$. At this point, we
need to change $F_1$ and $F_2$ so that we can re-interpret the system
as having positive $\bns$-flux. The required change is:
\eqn\reqchange{
\eqalign{
F_1 =0\quad &\rightarrow\quad F_1 =1\cr
F_2 =1\quad &\rightarrow\quad F_2 =2\cr }}
This allows us to claim that the $\bns$ flux is $(1-\epsilon)$, which
is positive and within the desired range. Notice that in the process,
$F_-$ is unchanged but $F_+$ has jumped. The total D3-brane number
is conserved, as it must be since the system initially consisted of
one D3-brane at the conifold.

Note that the jump in $F_+$ does not really correspond to the creation
of any object as we pass through this phase transition. The
world-volume action of the D5 anti-D5 pair contains the couplings:
\eqn\dantid{
\int\,\left(F_-\wedge D^+ + (\bns - F_+)\wedge {}^*(\bns - F_+) + 
F_- \wedge {}^*F_- \right)}
where the first term is a Chern-Simons coupling and the remaining ones
come from the Dirac-Born-Infeld action\foot{As pointed out in
Ref.\refs\piljin, for a brane anti-brane pair the terms which come
from the DBI action, unlike the Chern-Simons terms, have to be added
rather than subtracted, because they involve a scalar product rather
than a wedge product of tensors.}. The gauge transformation that we
performed leaves the whole expression above unchanged.

Thus we learn that the K\"ahler transition for the simple conifold can
be interpreted as a jump in the value of the total world-volume field
strength $F_+$ on the brane anti-brane pair. The situation will be
more interesting for the generalized A-D-E conifolds\refs\gjm,
considered in the present context in
Refs.\refs{\afhs,\mori,\angel,\gns,\vonunge}, where the brane
construction has several NS and NS' 5-branes and by passing them
through each other the ordering is changed. This should be worth
exploring in more detail.

\newsec{Domain Walls}

We have been discussing two related theories, one involving N
D3-branes at a $Z_2$ ALE singularity, and the other involving the same
branes but at the node of a conifold. We will now be interested in the
large-N limit of these two theories.

In this limit, the former system is dual to type IIB string theory
compactified on $AdS_5\otimes (S^5/Z_2)$, where $Z_2$ reverses 4 of
the 6 directions in which the $S^5$ is embedded. The latter, on the
other hand, is dual to type IIB theory on $AdS_5\otimes T_{1,1}$ where
$T_{1,1}$ is the Einstein manifold discussed in Ref.\refs\romans,
where it is obtained as a particular quotient of the form
$SU(2)\otimes SU(2)/U(1)$ preserving supersymmetry. In the present
context, the emergence of this space is a consequence of the fact that
the conifold is a fibration of a half-line over $T_{1,1}$. 

These two cases correspond to compactifications of string theory with
$\half$ and ${1\over 4}$-maximal supersymmetry respectively.  In both
cases, our goal is to understand what becomes of a fractional brane,
namely a D5-brane that was wrapped around the singularity, once we
introduce N D3-branes and go to large N. Since the considerations
are reasonably analogous for the two cases, we will only discuss the
second one in detail. For this, we will need some details of the
geometry of the space $T_{1,1}$.

The metric of $T_{1,1}$ can be expressed in terms of 5 angular
coordinates, $\psi,\theta_1, \phi_1,\theta_2,\phi_2$ as\refs\candelas:
\eqn\conifmetric{
ds^2 = 
{1\over 6} \sum_{i=1}^2 (d\theta_i^2 + sin^2\theta_i\, d\phi_i^2)  +
{1\over 9}
(d\psi + cos\theta_1 d\phi_1 + cos\theta_2 d\phi_2 )^2 }

$T_{1,1}$ is topologically just $S^2 \otimes S^3$. The $S^2$ here is
precisely the result of blowing up the vanishing 2-cycle in the
conifold. The blowing up is not the usual resolution of the
singularity, but rather a change in geometry induced by the distortion
of space in the neighbourhood of N D3-branes.

Hence we start with the type IIB theory on a conifold, with N
D3-branes located at the node. Wrap a D5-brane on the $S^2$ factor of
the base $T_{1,1,} = S^2 \otimes S^3$. This is the domain wall of
Ref.\refs\gubkleb.

As we will see below, this $S^2$ is really the difference (in the
sense of homology) of the two $S^2$'s which form the base of the
$U(1)$ fibration giving $T_{1,1}$. In other words, the the $S^2$ on
which the 5-brane is wrapped to give a domain wall can be written 
\eqn\diffstwo{
S^2 = (S^2)_1 - (S^2)_2}
In the brane construction, $(S^2)_1$ is just the compactified (45)
plane while $(S^2)_2$ is the compactified (89)
plane\refs\dmconif. This means that the dual object to the domain wall
in the type IIA picture is something that carries a charge away from
the (45) plane and deposits it on the (89) plane.

A 4-brane ending on the first NS 5-brane carries away a charge from it,
and if it also terminates on the NS' 5-brane then it deposits the
charge on that.  Moreover, since the boundary of a 4-brane ending on a
5-brane is like a 3-brane (in the (0123) directions), the transverse
space to the boundary is a 2-dimensional space, namely the (45) plane or
the (89) plane respectively. Hence such a 4-brane carries charge away
from the (45) plane and deposits it on the (89) plane.

This shows that the dual to the domain wall is a 4-brane stretched
between the NS and NS' 5-branes. It does not go all the way around the
$x^6$ direction, but stretches along only one of the two segments
between the branes. Thus, as promised, we see that the brane on an
interval (which we have identified above with a fractional brane) maps
onto an exotic domain wall formed by a D5-brane wrapping a cycle of
{\it finite} size, once we go to the near-horizon AdS description.

We can exploit the relationship between this model and the $\cN=1$
supersymmetric MQCD model of Ref.\witqcd\ to make this more precise. As
we pointed out in Section 3, a D4-brane stretched between an NS5 and
an NS5' brane becomes joined into a single M5-brane in the M-theory
limit, illustrated in Fig. (3.3). The configuration in this figure
resembles the brane construction of Ref.\refs\witqcd\ with the
difference that here the $x^6$ direction is compact (as a result, the
model with $n$ fractional branes has gauge group $SU(N+n)\otimes SU(N)$
with negative $\beta$-function for the first factor and positive
$\beta$-function for the second. For large $N$, these
$\beta$-functions will only show up to subleading order in $1/N$.)

Now we are interested in going to an $AdS$ limit, so we start with N
D3-branes at a conifold singularity and then wrap a D5-brane around
the $S^2$ cycle of $T_{1,1}$. In the brane construction, this
corresponds to N D4-branes wrapping all the way around $x^6$ and an
additional D4-brane stretching along one direction between the NS and
NS' 5-branes. In the M-theory limit of this configuration, the fully
wrapped D4-branes turn into N M5-branes toroidally wrapped on $x^6,
x^{10}$. These are decoupled from the remaining branes of the
problem. The single D4-brane on an interval, along with the NS and NS'
5-branes on which it ends, turns into a single M5-brane which was
extensively analyzed in Ref.\refs\witqcd. In our conventions, it is
appropriate to define $v=x^4 + ix^5$ and $w=x^8 + i x^9$ (these differ
from the conventions in Ref.\refs\witqcd\ by the interchange
$x^7\leftrightarrow x^9$). We will also need the variable $t =
\exp -(x^6 + i x^{10}/R)$, although this is not periodic in $x^6$
so we should use it only for a finite range of values of $x^6$. The
result of Ref.\refs\witqcd\ is that the single M5-brane in question is 
wrapped on a holomorphic curve $\Sigma$ in the complex $(v,w,t)$ space 
given by (recall that there is a single D4-brane for the moment, so
the parameter $n$ in Ref.\refs\witqcd\ is equal to 1):
\eqn\witeqn{
v = w^{-1} = t}
Recalling that the $v$ and $w$-planes are related\refs\dmconif\ to
what we have been calling $(S^2)_1$ and $(S^2)_2$ above, these
equations give a precise meaning to our earlier statement that the
fractional brane is an object wrapping $(S^2)_1 - (S^2)_2$ and
stretching along $x^6$.

{}From the brane construction, it is clear that the gauge group jumps
from $SU(N)\otimes SU(N)$ to $SU(N+1)\otimes SU(N)$ as we cross such a
domain wall, as predicted in Ref.\gubkleb\ from different
considerations. In our picture, the enhancement of the gauge group is
due to open strings connecting the D4-brane on the interval and the
remaining D4-branes to which it is parallel and coincident over a
segment. However, we will see shortly that the full story is more
interesting.

To complete the above arguments, we must show that $S^2$ of $T_{1,1}$
is $(S^2)_1 - (S^2)_2$. A basis of vielbeins on $T_{1,1}$ is given in
Appendix (A) of Ref.\refs\candelas. One can make various 2-forms out
of these, but the question is which ones are in the cohomology. In
particular, we can write the two 2-forms
\eqn\twotwo{
\sin\theta_1\, d\theta_1\, d\phi_1 \pm \sin\theta_2\, d\theta_2\, 
d\phi_2}
both of which live only on the two $S^2$ factors in $T_{1,1}$, and are
independent of the $U(1)$ fibre.

Both these 2-forms can be written formally as exact forms, namely the
above expressions are equal to
\eqn\formalexact{
d( \cos\theta_1\, d\phi_1  \pm  \cos\theta_2\, d\phi_2)}
However, the expressions in brackets above are ill-defined when any of
the $\theta_i$ is equal to 0 or $\pi$, since in that case we are at
the north or south pole of one of the 2-spheres and the coordinate
$\phi_i$ is undefined there. However, the term with the $+$ sign can
be modified to:
\eqn\modifto{
d( d\psi + \cos\theta_1\, d\phi_1  +  \cos\theta_2\, d\phi_2)
\sim d e^{\psi}}
where $e^{\psi}$ is one of the five vielbeins, and is globally defined
because $\psi$ is allowed to have gauge transformations. It follows
that the term with the $+$ sign in Eq.\twotwo\ is genuinely exact,
leaving the one with the minus sign as the representative of the
second cohomology. {}From this in turn one deduces that in homology,
the $S^2$ factor in $S^2\times S^3$ is the difference of the two
$S^2$'s parametrized by $\theta_1,\phi_1$ and $\theta_2,\phi_2$
respectively, or in our brane construction, by $x^4,x^5$ and $x^8,x^9$
respectively.

Returning now to the exotic domain wall, we have observed that it is
the large-N version of a fractional brane. At a $Z_2$ singularity or a 
simple conifold, a very natural configuration (the one studied in
previous sections) consisted of a pair of fractional branes. The
analogue in the large-N limit will be a 5-brane anti-5-brane pair
wrapped around the 2-cycle of the Einstein space $T_{1,1}$. 

In Ref.\refs\gubkleb\ it was observed that while a 5-brane wrapped on
$S^2$ is an exotic domain wall which increments the gauge group from
$SU(N)\otimes SU(N)$ to $SU(N+1)\otimes SU(N)$, an anti-5-brane
wrapped on the same cycle reduces the gauge group back to
$SU(N)\otimes SU(N)$. According to our picture, the way to understand
this is that the wrapped 5-brane is a fractional 3-brane, while the
wrapped anti-5-brane is a fractional anti-3-brane. The two can then
simply annihilate. But now from the analysis in the previous
sections, we see that there is a new possibility. Suppose that on the
wrapped 5-brane anti-5-brane pair we turn on a unit flux of the
relative world-volume gauge field $F_-$. The result is that the pair
annihilates into a D3-brane. On the gauge theory side, the gauge group
is incremented from $SU(N)\otimes SU(N)$ to $SU(N+1)\otimes SU(N+1)$.

This can be understood in two steps. Suppose we choose to turn on a
unit flux of $F_-$ by assigning $F_1=0,F_2=1$. Then the wrapped
5-brane increases the gauge group from $SU(N)\otimes SU(N)$ to
$SU(N+1)\otimes SU(N)$ while the second one further enhances it to
$SU(N+1)\otimes SU(N+1)$. We see that the difference between the two
factors of the gauge group is only partly a matter of convention. If
we choose conventions where a 5-brane (with no fluxes) increments the
first factor, then an anti-5-brane decreases the same factor. However,
as we will show directly below, a flux of $\int F_1 = m_1$ on the
5-brane decreases both factors by $m_1$ units, and a flux of $\int F_2
= m_2$ increases both factors by $m_2$ units. As a result, a 5-brane
with an arbitrary flux $m_1$ ends up changing the gauge group to
$SU(N+1-m_1)\otimes SU(N-m_1)$ while an anti-5-brane with an arbitrary
flux $m_2$ changes it to $SU(N-1+m_2)\otimes SU(N+m_2)$. In
particular, it follows that an anti-5-brane with one unit of flux
increments the second factor and not the first.

To show that world-volume fluxes have the claimed effect on the gauge
group of the theory, we use an observation in Ref.\refs\gubkleb.  The
baryon-like operators constructed there have quantum numbers $(N+1,1)$
and $(1,N+1)$ under the global $SU(2)\otimes SU(2)$ symmetry group of
the theory. This can be traced to the $(N+1)$-fold degeneracy of the
ground state for wrapped D3-branes (which classically are just
non-relativistic charged particles localized on $S^2$). This
degeneracy is attributed to the fact that 5-form flux through
$T_{1,1}$ gives rise to magnetic flux through $S^2$, which imposes a
minimum angular momentum on the charged particle propagating on
$S^2$. The world-volume fluxes in our picture are simply additional to
the magnetic flux coming from wrapped 5-branes. On a 5-brane, $m_1$
units of this flux decrease the degeneracy\foot{The degeneracy is
decreased and not increased because the world-volume gauge field
couples with the opposite sign from $\bns$.} from $N+1$ to $N-m_1+1$,
implying that the baryons have SU(2) quantum numbers equal to
$(N-m_1+1,1)$ and $(1, N-m_1+1)$. It follows that the world-volume
flux alone would convert the gauge group to $SU(N-m_1)\otimes
SU(N-m_1)$, which is what we wanted to show.

One may ask whether our picture of the relationship between
world-volume fluxes and the tachyon condensation phenomenon on the
brane anti-brane pair survives the large-N limit. The string theory on
$AdS_5 \otimes T_{1,1}$ is the large-N limit of N D3-branes at a
conifold, and we have argued in a previous section that for any N
there is no tachyon on the pair. However, in the large-N limit the
geometry of the spacetime is completely modified. One consequence is
that the exotic domain wall has a finite tension even in the absence
of $\bns$-fluxes and world-volume field strengths. This tension will
be of order the product of the D5-brane tension and the volume of
$S^2$, or roughly
\eqn\domtension{
{1\over g_s (\alpha')^3}\times R^2 \sim {N\over \sqrt{g_s N} 
(\alpha')^2}}
This contrasts with the fact that the corresponding object in flat
space would be tensionless. Since that fact was crucial in arguing for 
the absence of tachyons on the brane anti-brane pair, it remains to be 
understood what is the correct statement that one can make about
tachyons on a domain-wall anti-domain-wall pair.

The fact that the $\bns$ flux through $S^2$ changes the tension of the
domain wall can be seen quite directly. In this situation there is a
nonzero field strength of $H_{RR}=dB_{RR}$ through the $S^3$ factor on
one side of the domain wall. The $\bns$ flux creates a net 3-brane
charge, which is a flux of $dD^+$. Thus we have nonzero and correlated
values of $\bns$, $dB_{RR}$ and $dD^+$, as expected from the fact that
the spacetime type IIB theory has a coupling $\int
\bns \wedge dB_{RR}\wedge dD^+$.

\newsec{Other Extended Objects}

In the type IIB compactification on $AdS_5\otimes T_{1,1}$ there are
stable objects arising from wrapped branes over various cycles. For
example, a ``fat string'' arises\refs\gubkleb\ from wrapping a
D3-brane over the $S^2$ factor in $T_{1,1}$. This object can be easily 
identified in the T-dual brane construction. It is a 2-brane
stretching between the NS and NS' 5-branes. Thus it extends for a
finite distance along $x^6$ and for an infinite distance along one of
$x^1,x^2,x^3$ depending on how we choose to orient the string.

In the brane construction it is clear that if we stretch a D2-brane
and a D4-brane between an NS and NS' 5-brane then all supersymmetries
will be broken, in agreement with the fact that this fat string is not
expected to be a BPS object. However, from the fact that $S^2$ is a
nontrivial homology cycle, the fat string appears to be stable. It is
interesting to compare this string with the QCD string arising in the
model of Ref.\witqcd. Evidently the fat string is rather
different. The QCD string, while also non-BPS, can annihilate in
groups of $n$ where the MQCD gauge group is $SU(n)$. For $n=1$, the
case of a single fractional brane, the QCD string would therefore be
unstable. Moreover, even for arbitrary $n$, it is given by an M-theory
membrane wrapped in a way quite different from what we have described
above. For example, while our membrane stretches along $x^6$, the QCD
string arises from a membrane at a fixed value of $x^6$.

Besides the D3-brane and the D5-brane, one can also wrap NS5-branes or
more generally $(p,q)$ 5-branes on $S^2$. These are all obtained from
a D5-brane using S-duality. On the brane construction side, the
picture is clearest after going to the M-theory limit. In this limit,
we have a configuration as in Fig. (3.3). If we had started with a
D5-brane, the horizontal ``tube'' of M5-brane in the figure would have
been wrapped on the $x^{10}$ direction while extending along the $x^6$
direction. If instead we start with a $(p,q)$ 5-brane, then this tube
is wrapped on a $(p,q)$ cycle of the 2-torus whose coordinates are
$x^6,x^{10}$. 

Next, consider the D3-brane wrapped on $S^3$. Before taking the
large-N limit, this corresponds to the much-discussed BPS state which
becomes massless at the conifold point in the moduli space of a
Calabi-Yau manifold\refs\stromconif. At large-N, this state is no
longer massless, and instead corresponds to the ``baryon'' of
Ref.\refs\gubkleb. After T-dualizing, this state should be a 2-brane
wrapped on some combination of the $x^4,x^5$ and $x^8,x^9$
directions. This is because, as shown in Ref.\refs\dmconif, the
direction $\psi$ in the metric of $T_{1,1}$ (see Eq.\conifmetric) is
identified with the T-duality direction $x^6$. The cycle $S^3$ can be
viewed as a fibration of $\psi$ over a 2-sphere parametrized by some
combination of $\theta_1, \phi_1, \theta_2, \phi_2$. According to the
analysis of Ref.\refs\dmconif, these map to the $x^4,x^5,x^8,x^9$
directions. Thus, T-dualizing along the $x^6$ direction converts the
wrapped 3-brane to a 2-brane wrapped over a 2-cycle in these 4
directions.

In fact, as pointed out in Ref.\refs\dmconif, the $x^4,x^5$ and
$x^8,x^9$ pairs of directions should each correspond to 2-spheres
rather than planes as they do in the brane construction. This
limitation of the brane construction makes it somewhat difficult to
understand in more detail the state obtained by wrapping a 2-brane
over the 2-cycle above.

Finally, a D5-brane wrapping $S^3$ gives rise to a ``fat
membrane''. In the T-dual picture, this object will be similar to the
baryon above, except that it has two additional dimensions filling a
plane in the $x^1,x^2,x^3$ directions. Thus it is a D4-brane
stretching along, say, $x^1,x^2$ and a 2-cycle in
$x^4,x^5,x^8,x^9$. Although this object does not stretch along $x^6$,
it is wrapped on $x^{10}$. Hence again it will have $(p,q)$ duals
which are wrapped on a $(p,q)$-cycle of the $x^6, x^{10}$
2-torus. These will correspond in the type IIB theory to $(p,q)$ fat
membranes obtained by wrapping $(p,q)$ 5-branes over $S^3$.

It is intriguing that MQCD too has a membrane in it, the QCD domain
wall, described as an M5-brane wrapping a supersymmetric
3-cycle\refs\witqcd. The configuration above also lifts to an M5-brane
wrapping a 3-cycle, but it is not expected to be a BPS object, and the
wrapping directions are rather different. Nevertheless, we find it
curious that our model and MQCD both have some kind of fat string and
domain wall that are described respectively by an M-theory membrane
and a 5-brane. This may be related to the fact that our model in the
presence of exotic domain walls resembles MQCD, as noted in Section 4.

\newsec{Fractional Branes and Brane Creation}

In this section we will give a new argument for the existence of
fractional branes in conifold like models. The argument is based on
the technique of brane creation. Certain configurations of branes
suspended between two infinitely extended branes can be thought of as
being created by crossing the two extended branes. Since we are
looking for the T-dual of such models, it will be simpler to T-dualize
the {\it initial} configuration and then represent the crossing by
turning on a Wilson line in the final picture. This Wilson line will
give rise to configurations which will be interpreted as the T-dual of
the suspended brane. 

However, as we shall see, not all configurations of suspended branes
lead to fractional branes. The existence of fractional branes is due
to some special properties of Taub-NUT spaces which appear as a
consequence of T-duality on the infinitely extended branes. We start
with a configuration of a fundamental string between two D4-branes.

Consider two D4-branes along
\eqn\twodfour{
\matrix{D4:&~~&0&1&2&3&4&-&-&-&-&-\cr
D4':&~~&0&-&-&-&-&5&6&7&8&-\cr}}
where $x^9$ is a compact direction. Initially the two D4-branes are at
a finite distance along $x^9$.  When they cross on the circle, an
F-string is created. The reason is that a D4 is a magnetic source of
the 3-form potential $C_{\mu\nu\rho}$, and when another D4 crosses the
first one there is a jump in the flux of the corresponding field
strength $G_{\mu\nu\rho\sigma}$.  Due to the Chern-Simons coupling
$\int~G\wedge A$ on the world volume of the D4-brane ($A$ is the gauge
field) this jump leads to a coupling $\int A$ on the world-line of the
intersection point. This corresponds to a source term for the
world-volume gauge field, which will in general break
supersymmetry. To get a BPS state, according to Ref.\refs\calmal, we
have to excite another world-volume scalar, say $x^9$. Therefore when
one D4 crosses another, it pulls out a piece of the second D4 as a
``spike''. This spike is a fundamental
string\refs{\hanwit,\dfk,\bdm}.

We would like to determine the T-dual of this configuration, of two
D4-branes connected by a F-string. To do this we instead T-dualize the
initial configuration, of the two D4 before crossing, and then turn on
a Wilson line in the final picture. The T-duality is made along
$x^9$. The D4-branes will become two D5-branes intersecting along a
(wrapped) string.

The relative velocity of the D4-branes translates into a time-varying
Wilson line on the circle \refs\bdm:
\eqn\wilson{
\int d^2 x~ \left({\del x^9_{(1)}\over \del t} - 
{\del x^9_{(2)}\over \del t}\right)
~~~ \buildrel{T_9}\over \longrightarrow ~~~
\int d^2 x~ \left({\del A^9_{(1)}\over \del t} - 
{\del A^9_{(1)}\over \del t}\right)}
The RHS is a $1+1d$ chiral anomaly $\int\, \omega~ \epsilon^{ab}\,
\del_a A_b$, where $\omega$ is the gauge-transformation parameter. This
anomaly is due to a chiral fermion propagating along the intersection
of the two D5 branes\foot{Since this fermion at the intersection is
chiral, we cannot give it a mass by moving the two D5-branes
apart. That this is the case is apparent from the world-volume
directions of the two D5 branes or, in the T-dual, of the two
D4-branes. This shows that brane creation can only occur if we have
have branes which together fill out eight spatial directions}.

The term which cancels the anomaly is 
\eqn\anomterm{
S = \int\, H_{RR}\wedge A\wedge F}
on the world volume of each D5. $H_{RR}= dB_{RR}$, the pullback of the
background three form, in the absence of any source. The cancellation
takes place via anomaly inflow. We have a coupling, eq \anomterm\ , in
$5+1d$ spacetime. Along a $1+1d$ subspace of this, chiral fermions
propagate and give rise to the anomaly Eq.\wilson. Now in the original
picture, of two D4-branes crossing, we saw that on {\it each} D4-brane
there is a change of flux of $G$. Therefore here, since D5-branes are
a magnetic source of $H_{RR}$, we find that changing the Wilson line
produces a change of flux of $H_{RR}$. In other words, a gauge
transformation $\delta A = d\omega$ on the world-volume will vary
Eq.\anomterm\ by $-\int dH_{RR} \wedge (\omega F)$. Since $dH_{RR}
\ne 0$ in the presence of a magnetic source of $H_{RR}$ flux, we end
up with:
\eqn\anomcancel{
\delta S = - \int d^2 x~ \omega\, \epsilon^{ab}\del_a A_b}
This is the inflow which cancels the anomaly.

Thus in the T-dual picture we get 0-branes (which are chiral fermions
coming from the string joining the D5-branes) on the two
D5-branes. The anomaly is cancelled by inflow. To summarise, we see
that in the D4-D4 system the flux change creates a spike to preserve
supersymmetry, while in the D5-D5 system the flux change creates an
anomaly inflow to cancel the anomaly due to chiral fermions.

Now consider another example of a D4-brane between an NS5-brane and a
D6-brane. This configuration is closely related to the configurations
we have been studying in the previous sections. The orientations of the
branes are as follows: 
$$
\matrix{NS5:&~~&0&1&2&3&4&5&-&-&-&-\cr
D4:&~~&0&1&2&3&-&-&6&-&-&-\cr D6:&~~&0&1&2&3&-&-&-&7&8&9\cr} 
$$ 
We are considering this configuration because a D4-brane gets created
when we move a D6 across NS5 along $x^6$, which is chosen to be a
compact direction of radius $R$. We want to determine the resulting
configuration after performing a T-duality along $x^6$.

The idea, as before, is to T-dualize the initial configuration of an
NS5 and a D6-brane. The crossing of the branes will now be reflected
as an asymptotic Wilson line on the $7+1$ dimensional gauge theory.

The reason why a D4-brane is created when we move a D6 across an NS5
is as follows. On the world volume of the NS5-brane there propagates a
chiral (2,0) tensor multiplet whose fields are $(B^+_{\mu\nu},
5\phi)$. The NS5-brane and the D6-brane are magnetic sources of the
spacetime 2-form potential $B_{NS,NS}$ and 1-form RR potential
$A$ respectively. The relevant coupling on the world volume of
the NS5-brane is
\eqn\nsfive{
\int~ A \wedge *d\phi = \int~ A \wedge dC_4}
$dC_4 = *d\phi$ is the six dimensional dual of a world-volume scalar.
{}From the previous arguments we see that there is a source of $\int~
C_4$ on the world volume of the NS5 when it is crossed by a D6-brane.
By itself, such a $C_4$ background will break supersymmetry. To
preserve supersymmetry we need to excite a scalar $x^6$ which will
satisfy
\eqn\scalar{
\del^2 x^6 = \delta (x)\delta (y)}
where $x, y$ are coordinates of the world-volume. Observe that
$\del^2$ involves {\it all} the six coordinates of the NS5-brane, but
$\delta (x) \delta (y)$ depends only on two of these coordinates.
Therefore the spike will be translationally invariant along three
directions. This implies that we have a {\it four-brane}. A similar
conclusion can also be arrived at by doing a series of S and
T-dualities to the D4-D4 system.  The T-dualities are all done
orthogonal to the compact direction.

{}From the D6-brane side we have a coupling 
\eqn\dsix{
\int~ B_{NS,NS} \wedge *F = \int~B_{NS,NS} \wedge dC_4}
This clearly leads to the same result.

Now, as before, consider T-dualizing the initial configuration. The
NS5 brane will become a Taub-NUT space and the D6-brane will become a
D7-brane wrapping the Taub-NUT space (which has a non-trivial metric
along the $(x^6,x^7,x^8,x^9)$ directions). The multiplet propagating
on the ``Taub-NUT 5-plane'', i.e. along the
$(x^0,x^1,x^2,x^3,x^4,x^5)$ directions, is again $(B^+_{\mu\nu},
5\phi)$. The self-dual antisymmetric tensor $B^+_{\mu\nu}$ comes from
reducing the ten-dimensional RR 4-form $D^+_{\mu\nu\rho\sigma}$ as
$B^+_{\mu\nu}(x)\otimes L_2(y)$, where $x,y$ are coordinates along the
012345 directions and the Taub-NUT space respectively, and $L_2(y)$ is
the normalizable harmonic two-form on Taub-NUT space. The five
scalars can be identified as follows: two of them come from the
axion-dilaton of type IIB and another two arise as an $L_2$ reduction
of the NS-NS and RR B-fields of the type IIB theory. The fifth scalar
is the gravity fluctuation. Various couplings of these fields with
background IIB fields can be worked out easily starting from the
non-selfdual action of IIB supergravity.

The D7-brane wrapping the Taub-NUT space will give rise to a D3-brane
bound to it. The charge of the D3-brane is given by the non-trivial
$B_{NS,NS}$ background on the Taub-NUT. To see this, consider some of
the couplings on the world volume of the D7-brane (we are neglecting
constant factors in front of each terms):
\eqn\coupdseven{
\int *\tilde\phi ~+~ \int D^+ \wedge F \wedge B_{NS,NS} ~+~ 
\int  D^+ \wedge F\wedge F ~+~ ...}
These couplings are derived from the WZ coupling $\int C\wedge
e^{B-F}$, where $C$ is the formal sum of the RR potentials.  The first
term $\int *\tilde\phi$ gives the charge of the D7-brane.

Now the motion of the D6 will turn into a Wilson line on the D7. The
point where the two branes, NS5 and D6, touch is the zero of the
Wilson line. Any positive value of the Wilson line will tell us
how far apart the two branes are after crossing. However, observe that
there is {\it no} global cycle now. Far away from the centre of
the Taub-NUT, the space looks like $R^3\times S^1$ but there is no 
non-trivial circle at the centre. Therefore, we cannot turn on a flat
connection on this space. Instead, a self-dual connection can be
turned on\foot{We would like to thank Anton Kapustin and Angel Uranga for
discussions on this point.}. This self-dual connection satisfies the
following equation on the Taub-NUT space:
\eqn\selfdual{
F= dA = L_2}
where $L_2$ is the unique normalizable harmonic two-form on the
Taub-NUT space. This harmonic two form, being normalizable, goes to
zero at infinity, hence we have a flat connection there. At infinity
there is an $S^1$ and therefore the flat connection corresponds to
a Wilson line.

In such a background, we can decompose the field strength $F$ as
\eqn\fdecom{
F = L_2 + F_1}
$F_1$ will now appear as a gauge field on the D7 (or Taub-NUT plane).
Inserting Eq.\fdecom\ in Eq.\coupdseven\ and integrating out $L_2$, we
have the following couplings:
\eqn\coupfinal{
\eqalign{
&{\rm 7-brane:} \qquad\int *\tilde\phi + \int D^+ \wedge F_1\wedge F_1\cr 
&{\rm 5-brane:} \qquad\int D^+ \wedge (B_{NS,NS}-F_1)\cr
&{\rm 3-brane:} \qquad\int D^+\cr}}
The term $\int D^+$ is just the D3-brane charge. This is 
the usual D3 which appears as an instanton on the D7. In the original
picture this will correspond to a D4 stretching all the way around the 
compact circle $x^6$.

The 5-brane term is more interesting. This also gives rise to a
D3-brane charge. But the charge is measured by $\int
B_{NS,NS}$\foot{This is because, as discussed in section 3, we can
gauge away the world volume field strength $F_1$ by a $B$-field gauge
transformation}. Therefore we have here another source of D3-brane
charge -- a wrapped five brane. If $\int B_{NS,NS}$ is fractional we
will get a fractional D3-brane here. Since $B_{NS,NS}$ and $L_2$ are
defined only on the Taub-NUT space, terms like $\int C_{RR}\wedge F^m
\wedge B^n$ for $m+n > 4$ do not contribute to the result. This
fractional brane is related to the discussions of the previous
sections. We will make the identification more precise shortly.

Another point to consider is the following. In the original picture we
can have a situation in which the D6-brane is fixed but now the NS5
brane moves. In the T-dual picture for this case we will have the 
following sources of D3-brane charge:
\eqn\nsmoves{
\int D^+ \wedge B\wedge B + \int D^+\wedge B\wedge F}
The first measures the T-dual of the D4 around the circle and the 
second measures the D4 between between two branes. To see this observe 
that the role of $B$ and $F$ gets exchanged in (5.8). This is clear from
the fact that the quantity which is physically observable, i.e gauge
invariant, is $B-F$; and it measures the relative distance between 
the two branes, NS5 and D6.  

Now consider the asymptotic region in the Taub-NUT space (which is
being wrapped by a D7). We have a time varying Wilson line there. This
is the manifestation of the motion of a D6 on the circle $x^6$ in the
original picture.  Due to the varying Wilson line we have an {\it
apparent} anomaly, as before, in $1+1d$. To cancel this ``anomaly'' we
need the following WZ coupling
\eqn\wzcoup{
S = \int G_5 \wedge A_1 \wedge F_1}
where $G_5 = dD^+$ in the absence of any source. Under a local gauge
transformation $\delta A_1 = d\omega$ we get 
$$
\delta S = -\int \omega F_1
$$
This equation holds because $dG_5 \ne 0$ in the presence of a
D3-brane. The source in question is precisely the D3-brane whose
charge is measured by the non trivial $B_{NS,NS}$ background.

Therefore we see that the T-dual of a D4-brane between an NS5-brane
and a D6-brane is a D3-brane bound to a D7-brane and a Taub-NUT
space. The charge of the D3-brane is given by the non trivial
$B_{NS,NS}$ background on the Taub-NUT.

At this point we would like to give a consistency check for the
identification of the D3-brane charges made above. Consider a
situation in which, in the original picture, the D6-brane crosses the
NS5-brane on a circle many times. Every time it crosses the NS5, it
creates a new D4-brane between the two branes. For example, when it
crosses the second time there will be a new D4 {\it plus} the original
one which was created in the first crossing. This makes the total
count as one {\it complete} D4, i.e a D4 starting and ending on the
NS5, and two D4-branes stretched between NS5 and D6. Similarly the
third crossing will give a count of three complete D4-branes plus
three stretched D4's, and so on. Thus after $(n+1)$ crossings we find
a total of $1+2+3+....+n = {n(n+1)\over 2} $ complete
D4-branes\foot{Another way to see this is the following: Consider an
infinite sequence of NS5 branes along a {\it non compact} $x^6$
direction separated by a distance $R$. When a D6-brane moves along
$x^6$ crossing all the NS5-branes, we see that the number of D4-branes
increases linearly.}.  For large $n$, this grows as $n^2/2$. In the
T-dual picture, motion along $x^6$ will be replaced by a background $F
= nL_2$. {}From Eq.\coupdseven\ we see that this gives a D3-brane
charge of 
$$
{n^2\over 2}\int D^+
$$
confirming the identification that the third term in Eq.\coupdseven\
measures the T-dual of the D4 starting and ending on the NS5. In
addition to that we also have, from Eq.\coupdseven, another term which
goes as 
$$
n \int D^+\wedge B_{NS,NS}
$$
This is just the charge of the $n$ D4-branes, which, at the $n$th
crossing, were {\it between} the NS5 and D6. The number of such
D4-branes obviously grows as $n$.  Therefore the third term in
Eq.\coupdseven\ does indeed measure, in the T-dual, the charge of a D4
between the NS5 and a D6.

However in the process of creating many D4-branes by crossing the NS5
and D6-branes many times we should be careful not to violate the {\it
s-rule}\refs{\hanwit}. That this rule is respected can be seen from
the construction of multiple images. We have a situation in which
there is an infinite array of NS5-branes. The various D4-branes
created by crossing D6-branes now do not join the {\it same}
NS5-branes (although for the purpose of calculating the number of
complete D4-branes we have broken the D4-branes on the array of
NS5-branes). This is illustrated in Fig.(6.1). 
\bigskip

\centerline{\epsfbox{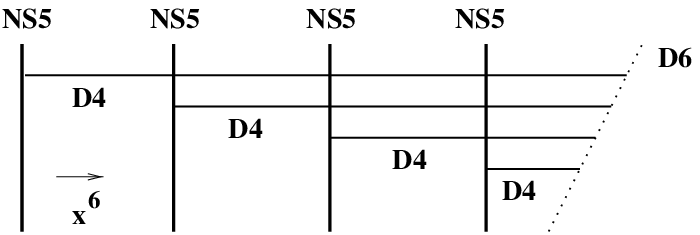}}\nobreak
\centerline{Fig. (6.1): D6-Brane Crossing NS5-branes.}
\bigskip

Now let us return to the conifold. As described in the previous
sections, what we want is the T-dual of a D4-brane between two
NS5-branes which are along the 12345 and 12389 directions
respectively. The T-dual of two intersecting NS5-branes gives a
conifold\refs{\angel,\dmconif}.

Since brane creation is a local process, the above analysis tells us
that the D4-brane becomes a D3 brane. We observed that in the previous
case the charge of the D3 was given by the $B_{NS,NS}$ background.
For the present case, the situation will be the same.  On the conifold
there can be a non-trivial $\bns$ background on the 2-cycle. Therefore
the D3-brane charge will now be given by
$\int_{S^2_{base}}~B_{NS,NS}$.  As discussed above, physically this is
just a D5-brane wrapped on the base of the conifold\foot{Another way
to see this is the following: For a bunch of D3-branes placed at the
conifold point we know that the dynamics of the D3-branes is governed
by an $\cN=1$, $SU(N)\otimes SU(N)\otimes U(1)$ gauge theory with
matter $A_i, B_i (i=1,2)$ in the antisymmetric representation of the
gauge group\refs{\klebwit,\angel,\dmconif} and with a quartic
superpotential. The D-term equation is given by $$D = |A_1|^2 +
|A_2|^2 - |B_1|^2 - |B_2|^2 - \zeta$$ where $\zeta$ is the coefficient
of the FI term in the $3+1d$ gauge theory. This $\zeta$ is also the
conifold flop factor\refs\klebwit. In the T-dual picture a flop
transition will be viewed as a motion of one of the NS5-brane along
$x^7$\refs\givkut. This motion will give mass to the D4-brane placed
between the two NS5-branes. Therefore, in the conifold model, as we
move the NS5 along $x^7$ a two cycle grows giving mass to the
D3-brane. This is only possible if the D3-brane was a wrapped
D5-brane. We thank Mike Douglas for discussions on this point.}.

For the case of two parallel NS5-branes the T-dual is an $A_1$
singularity. At the orbifold point, $\int B_{NS,NS}= {1\over
2}$. Therefore we have a fractional D3-brane whose charge is
half-integral.

Another interesting case where we can use brane creation to predict
the T-dual of some configuration is the original Hanany-Witten setup.
Consider the following brane configuration: 
$$
\matrix{NS5:&~~&0&1&2&3&4&5&-&-&-&-\cr
D5:&~~&0&1&2&-&-&-&-&7&8&9\cr
D3:&~~&0&1&2&-&-&-&6&-&-&-\cr}
$$
As before we assume the $x^6$ direction to be compact and the D3 to lie
between the two five branes. This configuration gives rise to $N=3, d=3$
gauge theory. 

If we take a configuration of an NS5 and a D5 with {\it no} D3-brane
between them then we can create a D3 by crossing the two branes.  The
analysis of brane creation for this configuration is identical to the
previous discussions. The NS5-brane is a source of
$B_{NS,NS}$. Therefore moving a NS5-brane across a D5 amounts to
changing the $B_{NS,NS}$ flux on the world volume of the D5 brane. Due
to the coupling $\int~B_{NS,NS}\wedge *F$ on the D5, we see that a
D3-brane gets created in the process. A similar argument can be given
for the case in which a D5 crosses an NS5-brane.

Now we T-dualize the initial configuration. The NS5-brane becomes a
Taub-NUT space and the D5 becomes a D6 completely wrapping the
Taub-NUT space. The motion of the branes in the original picture gets
replaced by an asymptotically varying Wilson line on D6 which goes
from negative to a positive value.

As before, we can analyse the world volume coupling on a D6 which is
wrapping a Taub-NUT space completely. Integrating out the $L_2$ we  have
the following sources of D2 brane charge:
\eqn\dtwo{
\int C + \int C \wedge (B_{NS,NS}-F_1)}
The first term is the usual T-dual of a D3 completely wrapping the
$x^6$ circle. The second term will give rise to D2-brane charge from
$\int B_{NSNS}$.  

Asymptotically, due to a varying Wilson line we have an apparent
anomaly. This is cancelled by the term $\int G_4 \wedge A_1 \wedge
F_1$, $G_4 =dC$ in the absence of any source. Therefore the T-dual of
a Hanany-Witten type configuration is a D2 brane bound to a D6 and a
Taub-NUT space. The charge of the D2 brane is given by the background
$B_{NS,NS}$ field.

We can lift this configuration to M-theory and try to see what the
brane creation process implies. A configuration of an NS5-brane and a
D5 brane intersecting on a $2+1$ dimensional space in type IIB theory
can be described as M-theory on a toric hyper-K\"ahler
manifold\refs\gnttown. On the IIB side, as we have seen before, one
can move the branes across each other to create a D3 between them. How
does one interpret this in M-theory?

The IIB configuration of an NS5 and a D5-brane on a circle T-dualizes
to the configuration, described above, of a Taub-NUT space and a
D6-brane. When we lift that configuration to M-theory, these just
become a pair of intersecting Taub-NUT spaces. This space has $Sp(2)$
holonomy and is a toric variety. The low energy dynamics of a {\it
single} Taub-NUT space is governed by a $U(1)$ gauge multiplet.
{}From the above analysis we found that the T-dual of a D3-brane
between two 5-branes was a D4-brane carrying a D2-brane charge.  In
M-theory we expect this to go to an M5-brane wrapped on a 3-cycle of
the hyper-K\"ahler manifold, and carrying an M2-brane charge.
\bigskip\medskip

\noindent{\bf Acknowledgements:} 
\bigskip

We would like to thank D.-E. Diaconescu, Mike Douglas, Anton Kapustin,
Govindan Rajesh, Savdeep Sethi, Shishir Sinha, David Tong, Sandip
Trivedi, Angel Uranga, Edward Witten and Zheng Yin for many valuable
discussions. We are particularly grateful to Ashoke Sen for sharing
several valuable insights with us. The work of KD is supported in part
by the U.S.  Dept. of Energy under Grant No. DE-FG02-90-ER40542.

\listrefs    
\end